# PARTICULATE MATTER DYNAMICS

Rodolfo G. Cionco, Nancy E. Quaranta* y Marta G. Caligaris

Grupo de Estudios Ambientales
Facultad Regional San Nicolás
Universidad Tecnológica Nacional
Colón 332 (2900), San Nicolás, Buenos Aires
e-mail: gcionco@frsn.utn.edu.ar, nquaranta@ frsn.utn.edu.ar, mcaligaris@frsn.utn.edu.ar
* Investigador CIC

**Palabras clave:** partículas gruesas, material particulado, modelos y simulaciones, contaminación atmosférica.

## Abstract

A substantial fraction of the particulate matter released into the atmosphere by industrial or natural processes corresponds to particles whose aerodynamic diameters are greater than 50 µm. It has been shown that, for these particles, the classical description of Gaussian plume diffusion processes, is inadequate to describe the transport and deposition. In this paper we present new results concerning the dispersion of coarse particulate matter. The simulations are done with our own code that uses the Bulirsch Stoer numerical integrator to calculate three-dimensional trajectories of particles released into the environment under very general conditions. Turbulent processes are simulated by the Langevin equation and weather conditions are modeled after stable (Monin-Obukhov length $L > 0$) and unstable conditions ($L < 0$). We present several case studies based on Monte Carlo simulations and discusses the effect of weather on the final deposition of these particles.

**Resumen**. *Una fracción importante del material particulado liberado a la atmósfera en procesos industriales o naturales corresponde a partículas cuyos diámetros aerodinámicos son mayores que 50 µm. Se ha mostrado para estas partículas que la clásica descripción de los procesos de difusión, tipo pluma gaussiana, es inadecuada para describir su transporte y deposición. En este trabajo se presentan nuevos resultados referentes a la dispersión de material particulado grueso. Las simulaciones se realizan con un código propio que utiliza el integrador numérico de Bulirsch y Stoer para calcular trayectorias tridimensionales de partículas liberadas al ambiente bajo condiciones muy generales. Los procesos turbulentos se simulan mediante la ecuación de Langevin y las condiciones meteorológicas se modelan según circunstancias de estabilidad (longitud de Monin-Obukhov $L > 0$) e inestabilidad ($L < 0$). Se presentan diversos casos estudiados en base a simulaciones tipo Monte Carlo y se discute el efecto de las condiciones meteorológicas sobre la deposición final de estas partículas.*

# 1. INTRODUCCIÓN

El estudio del material particulado atmosférico (PM), es uno de los temas de mayor actualidad e importancia de las ciencias medioambientales, particularmente en el estudio de subproductos industriales [1] y emisiones volcánicas [2]. PM es el nombre genérico que designa a partículas de tamaños variables (desde nanométricas hasta aproximadamente cien micrones de diámetro) y diferente composición, que se liberan a la atmósfera mediante diversos procesos. De estas partículas, aquellas cuyos diámetros *aerodinámicos* son menores que 10 μm (i.e., con velocidades de sedimentación iguales a las de una esfera de 10 μm de diámetro y 1 g cm$^{-3}$ de densidad) son las más nocivas para la salud, ya que permanecen más tiempo en suspensión y pueden llegar con facilidad hasta los alvéolos pulmonares [3]; por lo tanto, se han puesto los mayores esfuerzos en el modelado del transporte de esta clase de PM [4]. Sin embargo, las partículas con diámetros aerodinámicos > 10 μm, o *material particulado grueso* (PMc), también forma parte de los procesos generales de contaminación provocando, por ejemplo, trastornos del tracto respiratorio superior [5, 6]. En suma, el estudio de la dispersión y deposición de PMc es imprescindible para una descripción teórica completa del flujo de partículas contaminantes en un determinado sitio. Vesovic et ál. [7], han puntualizado la necesidad de mejorar la comprensión y la cuantificación del transporte y deposición de este tipo de material. Estos autores estudian PMc de diámetros entre 75 – 106,7 μm, comparando las concentraciones predichas por un clásico modelo de pluma gaussiana (Fugitive Dust Model), con un modelo desarrollado por ellos para resolver las ecuaciones de movimiento de las partículas liberadas en la atmósfera bajo condiciones de estabilidad neutral. Este modelo computacional está basado en un método de diferencias finitas con paso ajustable. El resultado de [7] es concluyente: el modelo gaussiano introduce sobreestimación en el pico de concentración y subestimación de la concentración de contaminantes corriente abajo, en forma estadísticamente significativa. En efecto, el PMc (que también podría denominarse PM *pesado*), está dominado por la gravitación y la interacción con la atmósfera mediante un arrastre de Stokes a bajos números de Reynolds [8], por lo tanto, no puede ser tratado como un fluido. Además Vesovic et ál., puntualizan que su código de dispersión es bidimensional debido al costo computacional que les supone evaluar las trayectorias en una tercera dimensión. Si bien la justificación para usar un código 2D es razonable debido a que las partículas básicamente van a seguir la dirección de la corriente principal del viento, la aproximación no es válida en situaciones donde los ejes del sistema de referencia utilizado no coincidan con la dirección del viento, ésta presente rotaciones o la intensidad de la turbulencia (siempre presente en todas direcciones) produzca sensibles apartamientos de las partículas respecto a la corriente principal. Más recientemente en [9], se ha revisado el tema confirmando los resultados de [7], concentrándose en "puffs" de baja altura (algunos metros), considerando condiciones meteorológicas más generales, remarcando la necesidad de profundizar el conocimiento de la dinámica atmosférica del PMc. En el Grupo de Estudios Ambientales de la Facultad Regional San Nicolás se viene trabajando en problemas de dispersión de este tipo de partículas, particularmente en el desarrollado de programas para el cálculo preciso y rápido de trayectorias tridimensionales de PMc que superen todas las limitaciones más arriba especificadas. Las primeras aplicaciones (sólo válidas para condiciones de estratificación neutral) y la validación de las rutinas fundamentales del programa que aquí se presenta se han reportado en [10]. En las páginas siguientes se describen las principales características del modelo implementado y el código de dispersión de PMc desarrollado, se presentan ejemplos concretos de aplicación con gran número de partículas liberadas a 25 m de altura en condiciones atmosféricas estables e inestables.

## 2. EL MODELO

Para determinar las trayectorias del PMc liberado en condiciones atmosféricas generales, deben integrarse sus ecuaciones de movimiento bajo la acción de la gravedad terrestre y la interacción con la atmósfera. Sea $\vec{r} = \vec{r}(t) = (x, y, z)$, la posición de una partícula en un sistema de referencia fijo en Tierra; salvo indicación en contrario se dirigirá el eje $X$ según la dirección horizontal coincidente con la corriente principal del viento; el eje $Z$ coincidente con la vertical del lugar y el eje $Y$ completando la terna directa; $\vec{v} = \dot{\vec{r}}(t)$ es la velocidad de la partícula; $\vec{u} = (u_x, u_y, u_z)$ es la velocidad del viento. Considerando las fuerzas actuantes (gravitación, flotación y arrastre de Stokes), la aplicación de la segunda ley de Newton provee

$$\frac{d^2 \vec{r}}{dt^2} = -\left(1 - \frac{\rho_l}{\rho_p}\right)\vec{g} + \frac{f}{\tau}(\vec{u} - \vec{v}), \tag{1}$$

donde $\rho_l$ es la densidad del fluido; $\rho_p$ es la densidad de la partícula; $\vec{g}$ es la aceleración de la gravedad; además

$$f = 1 + 0{,}15\, Re^{0{,}687} \tag{2}$$

es una función del número de Reynolds ($Re$) y proviene del régimen de Stokes para $Re$ bajos ($< 200$) que son los de interés para este trabajo [8, 11]; $\tau$ es el tiempo de relajación para una partícula en un régimen perfectamente laminar, definido como

$$\tau = \frac{m_p}{3\pi\mu D_p}, \tag{3}$$

donde $m_p$ es la masa de la partícula, $D_p$ es su diámetro y $\mu$ la viscosidad dinámica del fluido; $\tau$ es el tiempo que tarda la partícula en perder su memoria dinámica y responder ante cambios de velocidad en el fluido ideal.

Para caracterizar la velocidad del viento en la capa superficial y los perfiles de difusión turbulenta, se utilizó la teoría de semejanza desarrollada por Monin y Obukhov y la descripción lagrangiana correspondiente [8,12]. Las propiedades de esos flujos son expresadas como funciones de la velocidad de fricción y de la longitud de Obukhov $L$.

Para el caso de estabilidad neutral ($z/L = 0$) el viento se modela, como es usual, considerando un viento medio

$$\vec{u}_m = \frac{\vec{u}_*}{\kappa} \ln\left(\frac{z}{z_0}\right) \tag{4}$$

donde $\vec{u}_*$ es la velocidad de fricción o cizalla (con componentes $X$ e $Y$), $\kappa$ es la constante de von Kármán (usualmente 0,35 - 0,40) y $z_0$ la altura de la capa límite (definida a partir de las irregularidades del terreno). Para el caso general de estabilidad ($z/L > 0$)

$$\vec{u}_m = \frac{\vec{u}_*}{\kappa}\left[\ln\left(\frac{z}{z_0}\right) + 4{,}7\left(\frac{z}{L}\right)\right]. \tag{5}$$

Para el caso de estratificación inestable ($z/L < 0$)

$$\vec{u}_m = \frac{\vec{u}_*}{\kappa}\left[\ln\left(\frac{z}{z_0}\right) - 2\ln\left(\frac{1+x}{2}\right) - \ln\left(\frac{1+x^2}{2}\right) + 2\tan^{-1}x - \frac{\pi}{2}\right], \tag{6}$$

con $x = (1 - 5\,z/L)^{0{,}25}$. Las expresiones (5) y (6) convergen a (4) cuando $z/L \to 0$ (computacionalmente la condición neutral se alcanza cuando $z/L$ se hace menor que un valor adoptado a priori). Teniendo en cuenta una perspectiva lagrangiana, se han incluido fluctuaciones turbulentas a la velocidad ($\vec{u}_t$), representadas en sus tres componentes, como soluciones de la ecuación de Langevin [8, 12]

$$\vec{u}_t(t) = \vec{u}_t(t-dt)\exp\left(-\frac{dt}{\tau_L}\right) + \vec{\sigma}\ rnd\left(1 - \exp\left(-\frac{2dt}{\tau_L}\right)\right)^{1/2} \tag{7}$$

donde, $dt$ es el lapso de tiempo en el cual se evalúa la turbulencia; $rnd$ es un número aleatorio que se genera con distribución gaussiana en $(-1,1)$; $\tau_L$ es la escala de tiempo lagrangiana definida como

$$\tau_L = \frac{0{,}4\,z}{|\vec{u}_*|}\left(1 + \frac{5z}{L}\right)^{-1} \qquad \text{(caso estable)} \tag{8}$$

$$\tau_L = \frac{0{,}4\,z}{|\vec{u}_*|}\left(1 - 6\frac{z}{L}\right)^{1/4} \qquad \text{(caso inestable);} \tag{9}$$

$\vec{\sigma}$ es la dispersión de velocidades, definida a partir de un factor de la velocidad de fricción para casos estables y neutral

$$\vec{\sigma} = (2{,}4|\vec{u}_*|,\ 2{,}4|\vec{u}_*|,\ 1{,}25|\vec{u}_*|). \tag{10}$$

Para casos inestables, la dispersión de velocidades se define por

$$\sigma_z^2 = \sigma_x^2 = \sigma_y^2 = |\vec{u}_*|\left(2,26 - 6,6\frac{z}{L}\right)^{0,67}. \tag{11}$$

## 3. EL CÓDIGO

La Ec. (1) se integra mediante el método de Bulirsch y Stoer [13]. Para ello se realizó un programa ForTran que utiliza la rutina BSSTEP de Numerical Recipes [14]. El método usa extrapolación de Richardson para aproximar la función a integrar mediante una subdivisión automática y arbitrariamente pequeña del paso de integración máximo inicial ($h_1$), de tal forma que se satisfaga una tolerancia prefijada. Este proceso garantiza gran precisión con un costo computacional mínimo. El programa comienza leyendo un archivo de condiciones iniciales con posición y velocidad para una partícula, el usuario introduce un número semilla para la generación de la secuencia de pseudoaleatorios usados en la Ec. (7), los cuales se obtienen mediante un procedimiento de L´Ecuyer [14]. La Ec. (1) se descompone en un sistema de seis ecuaciones acopladas; en cada paso de integración se calculan las componentes deterministas de las fuerzas intervinientes y las estocásticas (turbulencia) evaluándose, simultáneamente, las condiciones aerodinámicas y meteorológicas respectivas. Las salidas dan posición y velocidad de las partículas cada cierto intervalo de tiempo fijado por el usuario, el tiempo físico de vuelo y las coordenadas y velocidades en el punto de impacto. Una vez que la partícula llega hasta la altura mínima de integración (ZMIN, generalmente = $z_0$) se imprimen las salidas y el programa reinicia, a partir de las mismas condiciones iniciales, generando para una partícula idéntica un ambiente turbulento distinto. De esta forma el código funciona mediante un procedimiento Monte Carlo que permite seguir un número arbitrario de partículas hasta su deposición final. El tiempo físico medio de vuelo (TVM) es un parámetro importante que el usuario debe evaluar, ya que evita la integración de trayectorias más allá del dominio lógico de interés. La validación del programa se hizo mediante diversas comparaciones con problemas balísticos totalmente integrables (viento con velocidad constante) y con experimentos de dispersión de esferas de vidrio liberadas desde alturas máximas de 15 m [7]. En esta última versión recientemente puesta al día, se modelan toda las condiciones meteorológicas a partir de la estimación de la longitud $L$ de Obukhov.

## 4. RESULTADOS

Primero, y a modo de ejemplo, se presenta la simulación de la dispersión de esferas de 106,7 µm de diámetro liberadas desde 15 m donde la velocidad del viento fue de 7,31 m s$^{-1}$; este es un caso emblemático de [7] que ha sido utilizado para contrastar resultados con experimentos de dispersión de esferas trazadoras de vidrio. La Fig. 1 (a) muestra las trayectorias (2D) de algunas partículas simuladas. A pesar de simularse un caso de estabilidad neutral (ausencia de variaciones verticales en la turbulencia) aquellas partículas que logran mayor dispersión en $z$ caen más lejos de la fuente. La Fig. 1 (b), muestra el correspondiente histograma de deposición. Las partículas se encuentran con mayor probabilidad alrededor de 125 m de la fuente; la asimetría clásica de las curvas de dispersión se evidencia por una mediana levemente mayor, 130 m, valor a partir del cual se cuentan igual número de partículas corriente arriba y corriente abajo según la dirección principal del viento.

En el presente trabajo el rango de interés está centrado en partículas de 50, 75 y 100 μm emitidas a alturas mayores que las consideradas en [7] y [9]; i. e., por una fuente a 25 m de altura, en condiciones atmosféricas estables, inestables y neutrales y corriente principal de viento unidireccional (para viento rotante, ver [10]). En primer lugar, es importante mencionar que se ha establecido mediante gran número de simulaciones, que las velocidades iniciales de emisión son poco importantes frente a la velocidad del viento, es decir, que efectos inerciales no son importantes para este tipo de partículas, las cuales quedan rápidamente sujetas a la interacción con el viento, para luego decantar gravitacionalmente. Además, efectos importantes de flotación sólo son observables para partículas de densidad menor que 2 g cm$^{-3}$.

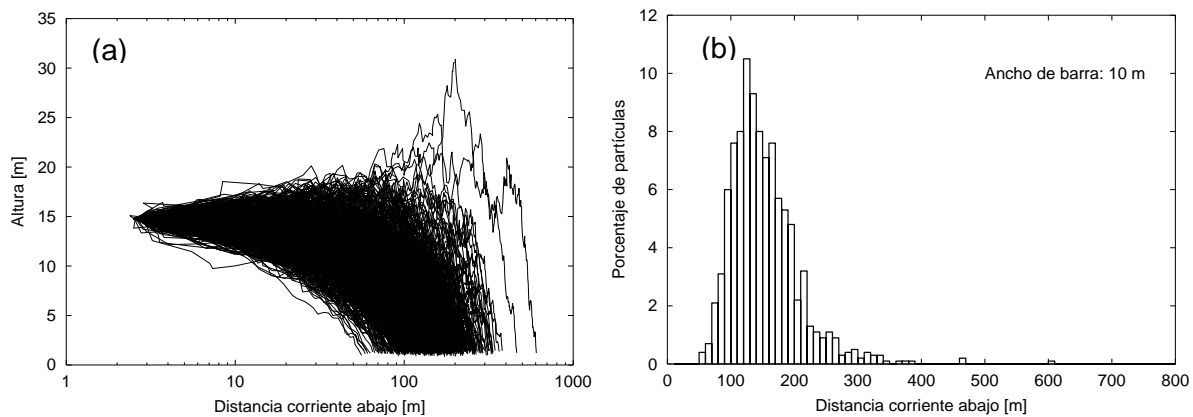

Fig. 1: (a) Trayectorias de $10^3$ partículas liberadas desde 15m. (b) Histograma correspondiente de deposición. El pico de la distribución está en la clase centrada en 125 m (ancho de clase 10 m); se encuentran partículas hasta una distancia de 605,4 m corriente abajo según la dirección del viento; a 200 m ha caído el 86,4 % del material particulado simulado.

A continuación se presentan algunos resultados de los casos simulados de PMc liberado a 25 m de altura bajo las tres clases de condiciones meteorológicas. El número de partículas utilizado fue de $10^4$ y se siguen hasta una altura ZMIN = 1,5 m; de acuerdo a esto (Ec. 8 y 9), el paso de integración $h_1$ debería ser del orden de 1 s; por seguridad, y para no perder resolución en posibles variaciones de la turbulencia, se ha usado un valor menor (0,4 s). La velocidad de fricción del viento utilizada fue de 1,6 km h$^{-1}$. En todos los casos se suponen partículas esféricas, con una densidad de 3 g cm$^{-3}$ similar al grafito. Con la finalidad de tener como referencia resultados publicados [9], aunque en un escenario ambiental diferente y con otra metodología de cálculo, se comenzó trabajando con longitudes de Obukhov de $\pm$ 200 m, para luego llegar a valores de $\pm$ 15 m. Para la simulación del caso neutral se fijó z /L = 0.

Para $L = \pm$ 200 m y el caso neutral correspondiente, las curvas de deposición no muestran grandes diferencias. Las mismas simulaciones con $L = \pm$ 15 m (que corresponden según [8] a condiciones estables-muy estables y muy inestables) muestran en cambio mayores discrepancias (Fig. 2). Para los casos estables, se verifica que el transporte de partículas aumenta sensiblemente a medida que $L$ disminuye. Este comportamiento se explica a partir del aumento de la velocidad de la corriente principal del viento (Ec. 5) y de la disminución con $L$ de la escala de tiempo lagrangiana (Ec. 8), lo cual refuerza el segundo término de la Ec. (7) permitiendo mayor dispersión en la velocidad turbulenta del viento. Este resultado es coincidente con el de Hubbard et ál. [9], quienes interpretan este fenómeno en términos de un aumento en los movimientos verticales en la masa de aire, lo cual produce mayor dispersión en Z, por lo tanto algunas partículas permanecen más tiempo en vuelo y son más fácilmente arrastradas por el viento. Este fenómeno puede observarse claramente en algunas trayectorias de la Fig. 1 (a). Los casos inestables ($L < 0$), no presentan variaciones tan marcadas, aunque el aumento de $L$ produce reducción en el transporte de PMc (el grueso de las partículas cae antes). Esto se debe a la disminución de la velocidad media del viento (Ec. 6); sin embargo, el aumento con $L$ del tiempo lagrangiano (Ec. 9), incrementa la velocidad turbulenta (refuerza el primer término de la Ec. 7) y la dispersión de velocidades del viento (Ec. 11); este último parámetro, mediante la aparición de valores *rnd* adecuados, puede tener un rol importante en la dispersión. En efecto, mientras en el caso $L = -15$ m el grueso de partículas cae antes que el correspondiente con $L = -200$ m, (como se observa fácilmente para las partículas de 50 μm de diámetro), una fracción apreciable cae más allá de los 2000 m (Fig. 2, b). Como es de esperar, el transporte más importante corresponde al de partículas de 50 μm las cuales al permanecer más tiempo en el aire, son más fácilmente arrastradas por el viento; por lo tanto, son más costosas de simular en términos de tiempo de cpu (*tcpu*). En general los tiempos de simulación varían en este caso entre 60 y 90 minutos, según sean las condiciones estables o inestables, para una computadora de 2,5 Ghz con recursos compartidos y Linux virtual bajo VmWare. Para estas partículas el TMV es de unos 200 s; sin embargo, en las simulaciones con $L < 0$ unas pocas partículas (entre 0,001 y 0,5 % para $L = -200$ m y -15 m respectivamente) fueron arrastradas a distancias mayores que 5000 m, que corresponde a tiempos físicos de vuelo mayores que 20 min; estos casos provocan que los tiempos de integración crezcan exponencialmente y que las partículas deban ser removidas de la simulación luego de ser reportadas en un archivo ad-hoc. Para la computadora mencionada y dentro del rango de tamaños de partículas involucradas, los tiempos de cpu varían linealmente con el inverso del diámetro de la partícula; por ejemplo, para el caso estable con $L = 200$ m se tiene: $tcpu = -48{,}94$ min $+ 5537{,}97$ min $(1\,\mu m / D_p)$.

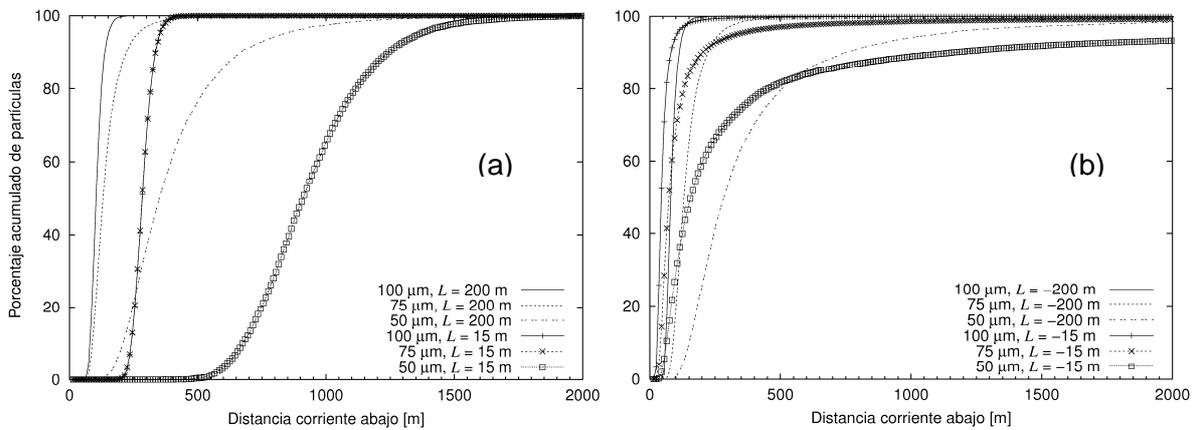

Fig. 2: (a) Porcentaje acumulado de partículas para casos estables con $L = 200$ m y $L = 15$ m; el transporte de partículas aumenta notablemente con $L$ menor. (b) Ídem para casos inestables con $L = -200$ m y $L = -15$ m; con $L$ mayores el grueso de las partículas cae antes pero aumenta la dispersión en la caída de partículas. Ambos fenómenos se deben a la variación de la velocidad media y turbulenta del viento. El caso neutral es bastante similar al inestable con $L = -200$ m.

Para una cuantificación adecuada de la deposición de partículas se calcula moda, mediana y distancia desde la fuente para la cual se encuentra depositado el 80 % del PMc simulado (X80) (Tablas 1 y 2). Como ya se mencionó, el transporte mayor corresponde al de partículas de 50 μm, estando los picos de la distribución sensiblemente desplazados respectos a los otros tipos de partículas simuladas.

|   |        | Moda ± 200 m | Mediana ± 200 m | X80 ± 200 m | Moda ± 15 m | Mediana ± 15 m | X80 ± 15 m |
|---|--------|--------------|-----------------|-------------|-------------|----------------|------------|
| E | 50 μm  | 184,88       | 372,83          | 535,50      | 875,32      | 915,90         | 1125,63    |
|   | 75 μm  | 115,74       | 130,62          | 180,95      | 285,44      | 280,12         | 316,16     |
|   | 100 μm | 128,88       | 108,52          | 135,37      | 295,61      | 283,39         | 315,93     |
| I | 50 μm  | 268,31       | 288,69          | 505,36      | 75,49       | 155,21         | 455,86     |
|   | 75 μm  | 158,45       | 135, 59         | 205,57      | 55,23       | 77,96          | 132,24     |
|   | 100 μm | 109,94       | 85,17           | 105,82      | 45,31       | 48,85          | 64,52      |

Tabla 1. Estadística relevante de las distribuciones de partículas obtenidas para el caso estable (E) e inestable (I), (con $L= \pm 200$ m y $L = \pm 15$); se consigna moda, mediana y distancia desde la fuente para la cual se deposita el 80 % del PMc simulado; unidades en metro.

|   |        | Moda   | Mediana | X80    |
|---|--------|--------|---------|--------|
| N | 50 μm  | 204,45 | 313,7   | 695,14 |
|   | 75 μm  | 137,02 | 149,32  | 215,78 |
|   | 100 μm | 103,81 | 94,02   | 115,76 |

Tabla 2. Estadística relevante de las distribuciones de partículas obtenidas para el caso neutral (N, $z/L = 0$); se consigna moda, mediana y distancia desde la fuente para la cual se deposita el 80 % del PMc simulado; unidades en metro.

Para los casos estables la estadística confirma el aumento de transporte y dispersión a medida que *L* disminuye. Para los casos inestables, la disminución de modas, medianas y del parámetro X80, a medida que *L* aumenta, confirma la reducción del transporte de PMc; la dispersión "anómala" de partículas más allá de los 2000 m no tiene incidencia en la estadística final, debido a que el porcentaje de partículas en esta situación no supera el 10 %.

Es interesante notar que en todos los casos, la turbulencia comienza a ser efectiva a partir de ~ 30 m de la fuente, i. e., para dos partículas idénticas bajo las mismas condiciones de simulación, sus trayectorias muestran una clara divergencia a partir de 30 m de la fuente. Por otra parte, por debajo de los 5 m de altura, la deposición de las partículas es manejada por asentamiento gravitacional, alcanzando las partículas velocidades del orden de 0,2 - 0,6 m s$^{-1}$.

## 5. CONCLUSIONES Y PERSPECTIVAS FUTURAS

En este trabajo se han reportado simulaciones de la dinámica de PMc liberado a la atmósfera, realizadas con un código propio. El integrador de Bulirsch y Stoer, de amplio uso en astrofísica, resultó ser sumamente preciso y rápido para este tipo de cálculos, a pesar de las dificultades inherentes a la turbulencia. En efecto, una de las limitaciones de este método es que la función a integrar sea suave [14]; sin embargo, no se encontraron problemas para simular la trayectoria de un gran número de partículas bajo todas las condiciones de estratificación atmosférica usuales. La subdivisión automática y arbitraria del paso de integración, hasta alcanzar la tolerancia deseada, evita problemas de convergencia relacionados con la longitud de $h_1$, el cual queda determinado por el tiempo lagrangiano mínimo para la turbulencia. Al respecto, la única limitación para la eficiencia del método es estimar previamente el tiempo físico de vuelo promedio para las partículas simuladas, debido a que algunas pueden permanecer en el aire durante lapsos de tiempo muy poco comunes, siendo arrastradas a distancias por fuera de la zona de interés o dominio lógico de integración. Las simulaciones de partículas entre 50 y 100 μm de diámetro, demandan casi 90 min para los casos más costosos, en una máquina modesta. Por lo tanto, el uso de este método, evita los inconvenientes computacionales reportados en la literatura para PMc pesado [7]. Las simulaciones realizadas indican que, a los efectos del transporte, las velocidades de emisión del PMc no son tan importantes frente a la velocidad del viento, parámetro que maneja el arrastre de las partículas. Para los casos estudiados bajo condiciones estables e inestables, el PMc se ve más fácilmente transportado corriente abajo a medida que la longitud de Obukhov disminuye. Este comportamiento se explica debido al aumento en la velocidad media del viento y también de la turbulencia, parámetro fuertemente dependiente de la escala de tiempo lagrangiana. Físicamente, $\tau_L$ maneja el tiempo para el cual las velocidades de las partículas del fluido cambian y se adaptan a las variaciones de la turbulencia. Por lo tanto, para un dado caso, $\tau_L$ menores favorecen la turbulencia en el fluido y una afectación mayor de las velocidades de las partículas inmersas en él.

Para una fuente situada a 25 m de altura, en una zona suburbana, con viento de 1,6 m s$^{-1}$ de velocidad en la capa límite, partículas mayores que 50 μm son depositadas casi en su totalidad dentro de los 1000 m corriente abajo; partículas de 50 μm se tiene mayoritariamente dentro de esta zona pero pueden encontrarse hasta unos 10000 m de la fuente. Estas conclusiones están apoyadas, además, por simulaciones con valores de *L* intermedios a los aquí reportados y muestran que este tipo de partículas presentan una dinámica compleja que es necesario continuar estudiando [7, 9]. Se espera de aquí en más continuar explorando el espacio de parámetros del problema y diversificar los casos de aplicación; en particular, adosar un término de convección-difusión para describir la dinámica de partículas de tamaños menores.

**REFERENCIAS**